\documentclass[a4paper,fleqn]{cas-sc}
\usepackage{lineno}
\usepackage[numbers,sort&compress]{natbib}
\usepackage{amsmath}
\usepackage{amssymb}
\usepackage{graphicx}
\usepackage{float}
\usepackage[section]{placeins}

\graphicspath{{figs/}}

\begin{document}
\let\WriteBookmarks\relax
\def\floatpagepagefraction{1}
\def\textpagefraction{.001}

\shorttitle{CNN-based position reconstruction for the T0 detector}
\shortauthors{Lv et~al.}

\title[mode=title]{CNN-based Position Reconstruction for the T0 Detector in the HIRFL-CSR External-target Experiment}
\tnotemark[1]

\tnotetext[1]{This work was supported by the National Natural Science Foundation of China under Grant Nos. 11927901, 12205296 and 11975228, the State Key Laboratory of Particle Detection and Electronics under Grant No. SKLPDE-ZZ-202416, and the USTC Research Funds of the Double First-Class Initiative under Grant No. WK2030000052.}

\author[1]{Shuai-Ke Lv}[style=chinese]
\fnmark[1]
\credit{Data curation, Writing - Original draft preparation}

\author[1]{D Hu}[type=editor,
                        auid=000,bioid=1,
                        orcid=0000-0001-0000-0000]
\cormark[1]

\ead{hurongr@ustc.edu.cn}

\credit{Conceptualization of this study, Methodology, Software}

\affiliation[1]{organization={Department of Modern Physics, University of Science and Technology of China},
                addressline={96 Jinzhai Road}, 
                city={Hefei},
                postcode={230026}, 
                state={AnHui},
                country={China}}

\author[1]{M Shao}[type=editor,
                        auid=000,bioid=1,
                        orcid=0000-0002-2268-5624]
\credit{Data curation, Writing - Original draft preparation}
\cormark[2]
\ead{swing@ustc.edu.cn}



\author[1]{Yingjie Zhou}
\credit{Data curation, Writing - Original draft preparation}

\author[1]{Yiding Zhao}
\credit{Data curation, Writing - Original draft preparation}


\cortext[cor1]{Corresponding author}
\cortext[cor2]{Principal corresponding author}
\fntext[fn1]{This is the first author.}


\begin{abstract}
The T0 detector provides the event start time for the HIRFL-CSR external-target experiment, and its multi-channel readout signals also contain two-dimensional position information of incident particles. In the CEE experiment, the dedicated beam-profile detector has a rate capability of only about 1 kHz, far below the beam event rate of up to 1 MHz, and therefore cannot provide real-time beam-spot monitoring under high-rate data-taking conditions or an event time reference. To fill this gap, this work investigates a convolutional neural network (CNN)-based position reconstruction method that enables the T0 detector to provide both the start time and beam-spot information. The over-threshold times and time-over-threshold (TOT) values from 32 SiPM channels are reorganized into a geometry-preserving input tensor. The time features are corrected by event-wise mean-time subtraction, while the TOT features are normalized by an event-wise trimmed mean before training-set feature standardization. A customized CNN then directly regresses the incident coordinates. Training and validation samples are constructed using Geant4 simulation combined with waveform digitization, and the method is experimentally evaluated with a pulsed-laser scanning data set with known position labels. The standard deviations of the residual distributions on independent test sets are approximately 1.11 mm and 1.04 mm in the x and y directions for uniformly distributed planar simulation data. For the discrete laser-scanning geometry, the corresponding resolutions are about 2.36 mm and 1.34 mm in simulation and 2.42 mm and 1.41 mm in experiment. The trained model is further applied to CEE zero-field beam data collected at the Institute of Modern Physics, Lanzhou, in January 2026, and a two-dimensional beam-spot distribution is reconstructed. These results demonstrate the potential of the T0 detector as a high-rate beam-spot monitor while retaining its role as an event start-time detector.
\end{abstract}

\begin{graphicalabstract}
\includegraphics{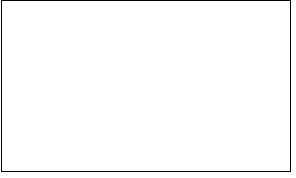}
\end{graphicalabstract}

\begin{highlights}
\item A CNN-based position reconstruction method is developed for the HIRFL-CSR T0 detector.
\item The network uses relative over-threshold time and event-normalized TOT information from 32 SiPM channels.
\item Simulation and laser-scanning data show millimeter-scale two-dimensional positioning performance.
\item A beam-spot distribution is reconstructed from CEE zero-field beam data collected at IMP Lanzhou.
\end{highlights}

\begin{keywords}
T0 detector \sep HIRFL-CSR \sep Position reconstruction \sep Convolutional neural network \sep SiPM \sep Geant4 \sep Laser scanning
\end{keywords}

\maketitle

\section{Introduction}

The HIRFL-CSR external-target experiment is designed for studies of particle production, transport processes, and properties of nuclear matter in intermediate- and high-energy heavy-ion collisions\cite{csrext,luo2017nst}. For such experiments, an accurate event start time and stable monitoring of the beam condition are essential for subsequent time-of-flight measurements, trigger selection, and event reconstruction. The T0 detector is located upstream of the target along the beam line and provides the event reference time using a fast plastic scintillator read out by SiPM arrays\cite{hu2023t0}. Because the detector uses multi-channel readout, different incident positions lead to variations in signal arrival time, signal width, and collected light among channels. Therefore, in addition to providing the event start time, the T0 detector also has a two-dimensional position response.

In the CEE experiment, beam-profile monitoring currently relies on an independent beam-spot positioning detector. However, its count-rate capability is only about 1 kHz, whereas the designed beam event rate can reach the order of 1 MHz. Because of this count-rate bottleneck, the detector cannot reflect the beam-spot status in real time under normal data-taking conditions, and it does not have an event start-time measurement function. Therefore, a beam-spot monitoring method with both high-count-rate tolerance and time-zero measurement capability is urgently needed. The T0 detector uses a fast scintillator and SiPM arrays, and its readout electronics (NINO + TDC) support event rates from hundreds of kHz to the MHz level. If the two-dimensional incident position can be reconstructed from its multi-channel response, high-count-rate online beam-profile monitoring can be realized while providing an accurate time reference, which is of key importance for beam diagnostics and offline data analysis in the CEE experiment.

Existing position reconstruction methods mostly rely on channel geometric centroids, time differences between two ends, or empirical calibration functions. However, for the T0 detector studied in this work, conventional methods face severe challenges. Light propagation in the scintillator includes boundary reflections and path losses, and a simple exponential attenuation model cannot accurately describe the actual light distribution. The SiPM gain is easily affected by temperature and bias voltage. The time walk introduced by threshold discrimination and the amplitude effect are nonlinearly coupled. In addition, the discriminator threshold of the NINO chip makes the relation between TOT and amplitude nonlinear. These factors make it difficult for analytical models to achieve global accuracy, and their generalization in different incident regions is poor, limiting further improvement of the positioning performance.

Deep learning, especially convolutional neural networks (CNNs), has shown strong nonlinear mapping capability in image recognition and particle-physics data reconstruction. For detector arrays, CNNs are naturally suitable for processing multi-channel signals with spatial ordering. In recent years, CNNs have been widely applied to calorimeter energy reconstruction \cite{alimena2020,polson2022energycnn} and to particle identification and track reconstruction in tracking detectors \cite {atlas2019cnn,microboone2021cnnpid}, but their applications to position reconstruction in T0/start detectors remain limited. Unlike fully connected networks that regard signals as independent features, the customized CNN in this work explicitly preserves the arrangement order of the SiPMs along the short axis on each side and the correspondence between the arrays at the two ends along the long axis, allowing the convolution kernels to effectively extract spatio-temporal correlation patterns along the light-propagation paths.

The main contributions of this work are as follows: (1) for the specific readout structure and channel arrangement of the T0 detector, an input-tensor construction method jointly encoding relative time and event-normalized TOT is designed, enabling the network to learn both short-axis neighborhood information and long-axis global response; (2) through Geant4 simulation and laser-scanning experiments, the performance difference of the CNN method under ideal continuous labels and practical discrete labels is systematically evaluated, and the effect of scanning-point density on supervised-learning accuracy is verified; (3) a CNN algorithm is used for the first time in this type of start-time detector to reconstruct the two-dimensional position, and the feasibility of the T0 detector to simultaneously provide time zero and beam-spot monitoring under high-count-rate CEE beam conditions is demonstrated.


\section{T0 detector}
\label{sec:detector}

\subsection{Detector structure}

The T0 detector studied in this work uses EJ-200 plastic scintillator\cite{ej200} as the sensitive medium. EJ-200 has a fast scintillation decay time, a long optical attenuation length, and good radiation tolerance, making it suitable for fast timing measurements in a heavy-ion beam environment. The active area of the detector is approximately $200~\mathrm{mm} \times 100~\mathrm{mm}$. The scintillator thickness is selected on the millimeter scale according to the ion species, counting rate, timing requirement, and material budget of the experiment, so as to balance detection efficiency, energy-loss control, and timing response.

A Cartesian coordinate system is used to describe the incident position in the detector plane. Taking the center of the detector as the origin. The direction along the long side of the scintillator is defined as the $x$ axis, while the direction along the short side is defined as the $y$ axis. Therefore, all $x$-direction results in the following correspond to the long-axis direction of the detector, and all $y$-direction results correspond to the short-axis direction.

The scintillation light is read out by two SiPM arrays located at the two ends of the detector along the long axis. The arrays use Hamamatsu S13360-3025PE SiPMs\cite{s13360}. Each array is arranged along the short-axis $y$ direction, which is therefore also the spatial ordering direction of the SiPM channels on one side. The present data and algorithm use 32 readout channels, corresponding to 16 SiPM channels on each side. Along the $y$ direction, the local incident position changes the collected-light fraction and response timing of neighboring SiPM channels. Along the $x$ direction, the optical path lengths to the arrays at the two ends are different, causing the relative timing and TOT distributions of the two sides to vary with position. Thus, the two-dimensional position information of the T0 detector is encoded in the joint response pattern of the 32 channels.

\subsection{Readout electronics}

Each SiPM array is connected to a front-end electronics (FEE) board. The FEE of the T0 detector uses the NINO chip to amplify and discriminate the fast SiPM pulses, converting analog pulses into digital pulses that can be recorded by the back-end electronics\cite{cao2022icetis}. A high-precision TDC records the over-threshold time of the digital pulse, while the time width between the leading and trailing edges gives the time-over-threshold (TOT). Therefore, the T0 readout chain does not store the full waveform for each event, but provides the over-threshold time and TOT for each channel.

This readout scheme directly corresponds to the CNN input features used in this work. The over-threshold time reflects the relative arrival time at which scintillation light reaches each SiPM channel and causes the signal to exceed the threshold, and therefore contains information on the light propagation path and incident position. The TOT is related to the pulse width, signal amplitude, and collected light, and can be used as a measure of the channel response strength. It also partly complements the amplitude-dependent variation in the leading-edge time. 

For position reconstruction, the readout system must preserve the consistency of the inter-channel timing and TOT information. Variations in channel gain, threshold settings, NINO discrimination characteristics, and TDC timing precision may introduce systematic differences in the measured features. Therefore, appropriate data preprocessing is required to reduce the influence of such effects and improve the consistency of the input data.
\section{CNN-based position reconstruction method}
\label{sec:method}

A convolutional neural network (CNN) is a class of neural network that extracts spatially correlated features through local connections, weight sharing, and multilayer nonlinear transformations. Compared with a fully connected network, a CNN does not need to learn an independent set of weights for every input variable. Instead, trainable convolution kernels slide over the input tensor and extract local patterns that appear repeatedly in adjacent regions. This property is well suited to detector-array signals with a fixed geometrical order.

For the $l$-th convolutional layer, let $a_i^{(l-1)}$ denote the $i$-th input feature map from the previous layer, $W_{ij}^{(l)}$ the convolution kernel connecting input feature map $i$ to output feature map $j$, and $b_j^{(l)}$ the bias term. The forward propagation of a two-dimensional convolution can be written as\cite{goodfellow2016}
\begin{equation}
z_j^{(l)}(u,v) =
\sum_i \sum_m \sum_n W_{ij}^{(l)}(m,n)
a_i^{(l-1)}(u+m,v+n) + b_j^{(l)}, \qquad
a_j^{(l)}(u,v)=f\left[z_j^{(l)}(u,v)\right].
\label{eq:conv}
\end{equation}
Here, $(u,v)$ is the position in the output feature map, $(m,n)$ is the index inside the convolution kernel, and $f$ is the nonlinear activation function. The network in this work uses the ReLU activation function, $f(z)=\max(0,z)$. ReLU suppresses negative responses while preserving the linear variation of positive responses, enabling the network to learn nonlinear detector responses while maintaining efficient training.

Batch normalization and pooling operations are introduced after the convolutional layers. Batch normalization can be written as\cite{batchnorm}
\begin{equation}
\hat{x}=\frac{x-\mu_B}{\sqrt{\sigma_B^2+\epsilon}}, \qquad
y=\gamma \hat{x}+\beta ,
\label{eq:bn}
\end{equation}
where $\mu_B$ and $\sigma_B^2$ are the mean and variance of a mini-batch, and $\gamma$ and $\beta$ are trainable parameters. Pooling compresses local regions and reduces the sensitivity of the model to statistical fluctuations and local abnormal responses.

In supervised learning, the CNN predicts coordinates through forward propagation, and the difference between the predicted and true coordinates is measured by a loss function. For the two-dimensional position regression problem, the mean squared error is used as the training objective:
\begin{equation}
L=\frac{1}{N}\sum_k
\left[
\left(x_{\mathrm{pred},k}-x_{\mathrm{true},k}\right)^2+
\left(y_{\mathrm{pred},k}-y_{\mathrm{true},k}\right)^2
\right].
\label{eq:loss}
\end{equation}
During back propagation, the gradients of the loss function with respect to the convolution kernels, batch-normalization parameters, and fully connected weights are calculated layer by layer. The optimizer then updates the network parameters. After multiple training epochs, the network learns the mapping from multi-channel input features to the two-dimensional incident position.

For position reconstruction in the T0 detector, the advantage of a CNN is that it can simultaneously use the spatial correlations among multi-channel signals. The SiPM channels on one side are arranged in a fixed order along the short-axis $y$ direction, and the timing and TOT responses of adjacent channels are not independent. Instead, they jointly reflect the incident position, light propagation path, and light collection efficiency. The two arrays are located at the two ends of the detector along the long-axis $x$ direction and therefore also carry information related to the difference in light propagation path. In this work, the time and TOT information of the 32 channels is reorganized into a tensor with spatial order, allowing convolution kernels to extract local response features along the channel array. The final fully connected regression layers output the two-dimensional coordinates. Thus, the detector signals are not treated simply as an unordered set of variables; the geometrical correlations in the channel layout are explicitly used.

\subsection{Input features and preprocessing}

The CNN input features for a single event consist of the over-threshold times from 32 channels and the TOT values from the same 32 channels, giving 64 input variables in total. The over-threshold time is the channel leading-edge time recorded by the high-precision TDC after FEE discrimination, and it reflects the relative timing at which scintillation light reaches different SiPMs and produces a discriminated signal. The TOT is the time width for which the same digital pulse stays above threshold, and it contains information on signal amplitude, collected light, and pulse shape. These two types of features are complementary for position reconstruction: the timing features are sensitive to the propagation-path difference between the two ends and to the response order of neighboring channels along the short axis, while the TOT features provide information on the collected-light fraction and signal strength of different channels.

For the T0 detector, the detector itself is used to define the event start time in the experiment. In the single-detector positioning data used here, there is no independent external time reference that can be used as a common zero time for all channel over-threshold times. Therefore, the absolute over-threshold times recorded by the TDC contain an event-level common offset that is unrelated to the incident position. If the raw absolute times are used directly as network inputs, the model is affected by an unstable time origin and cannot learn a robust position mapping.

To solve this problem, the multi-channel timing information of the T0 detector is used to define an event-internal reference time. For an event with valid channel set $S$ and number of valid channels $N_{\mathrm{valid}}$, the event mean time and relative time are defined as
\begin{equation}
t_{\mathrm{mean}}=\frac{1}{N_{\mathrm{valid}}}\sum_{i\in S}t_i,
\qquad
t_i^{\mathrm{rel}}=t_i-t_{\mathrm{mean}} .
\label{eq:mean_time}
\end{equation}
After this treatment, each event uses its own mean time as the relative time zero, and the network input is the pattern of channel times relative to this internal reference. The information needed for position reconstruction is contained in the inter-channel relative over-threshold time distribution. Without this treatment, the network lacks a stable time reference and the position mapping cannot be established reliably.

The TOT values are not subjected to event-mean time subtraction, but they are also not used as raw widths directly. Their role is to provide spatial distribution information related to collected light and to help the network learn nonlinear variations of the over-threshold time caused by pulse-amplitude effects. To reduce the influence of event-to-event variations in total light yield, global gain, or laser/beam intensity, the TOT values are normalized within each event. Specifically, the 32 TOT values in one event are sorted, the lowest and highest \texttt{trim\_count} values are discarded, and the mean of the remaining values is used as the event-wise TOT scale factor. Each channel TOT is then divided by this scale factor to obtain an event-normalized relative TOT.

Let the sorted TOT values be $\mathrm{TOT}_{(1)} \leq \mathrm{TOT}_{(2)} \leq \cdots \leq \mathrm{TOT}_{(32)}$, and let $m$ denote \texttt{trim\_count}. The event-wise TOT scale factor and normalized TOT are defined as
\begin{equation}
s_{\mathrm{TOT}}=
\frac{1}{32-2m}\sum_{j=m+1}^{32-m}\mathrm{TOT}_{(j)},
\qquad
\mathrm{TOT}_i^{\mathrm{norm}}=
\frac{\mathrm{TOT}_i}{s_{\mathrm{TOT}}}.
\label{eq:tot_norm}
\end{equation}
This trimmed-mean normalization preserves the relative distribution of TOT among channels while suppressing event-level scale fluctuations that are not directly related to the incident position. The features are then rearranged according to the channel order along the short axis on one side and the correspondence between the arrays at the two ends, so that adjacent input elements are as close as possible to the actual spatial adjacency. The relative time and event-normalized TOT features are finally standardized after being combined, allowing features with different units and numerical ranges to be trained stably in the same network.

The input vector for a single event can be written as
\begin{equation}
\mathbf{F} =
\left[t^{\mathrm{rel}}_1,\ldots,t^{\mathrm{rel}}_{32},
\mathrm{TOT}_1^{\mathrm{norm}},\ldots,\mathrm{TOT}_{32}^{\mathrm{norm}}\right].
\end{equation}
Before training, each feature dimension is standardized using the statistics of the training set, $F_k^{\mathrm{std}}=(F_k-\mu_k)/\sigma_k$, where $\mu_k$ and $\sigma_k$ are the mean and standard deviation of the $k$-th feature in the training set.

The supervised-learning label is the two-dimensional incident position $(x,y)$, in units of mm. Here, $x$ denotes the long-axis direction and $y$ denotes the short-axis direction of the detector. For the simulation data, the label is the true incident position specified in Geant4. For the experimental data, the label is the hole position of the laser-scanning plate. The training objective is to minimize the mean squared error between the predicted and true coordinates.

\subsection{Network architecture}

\begin{figure}
  \centering
  \includegraphics[width=0.9\linewidth]{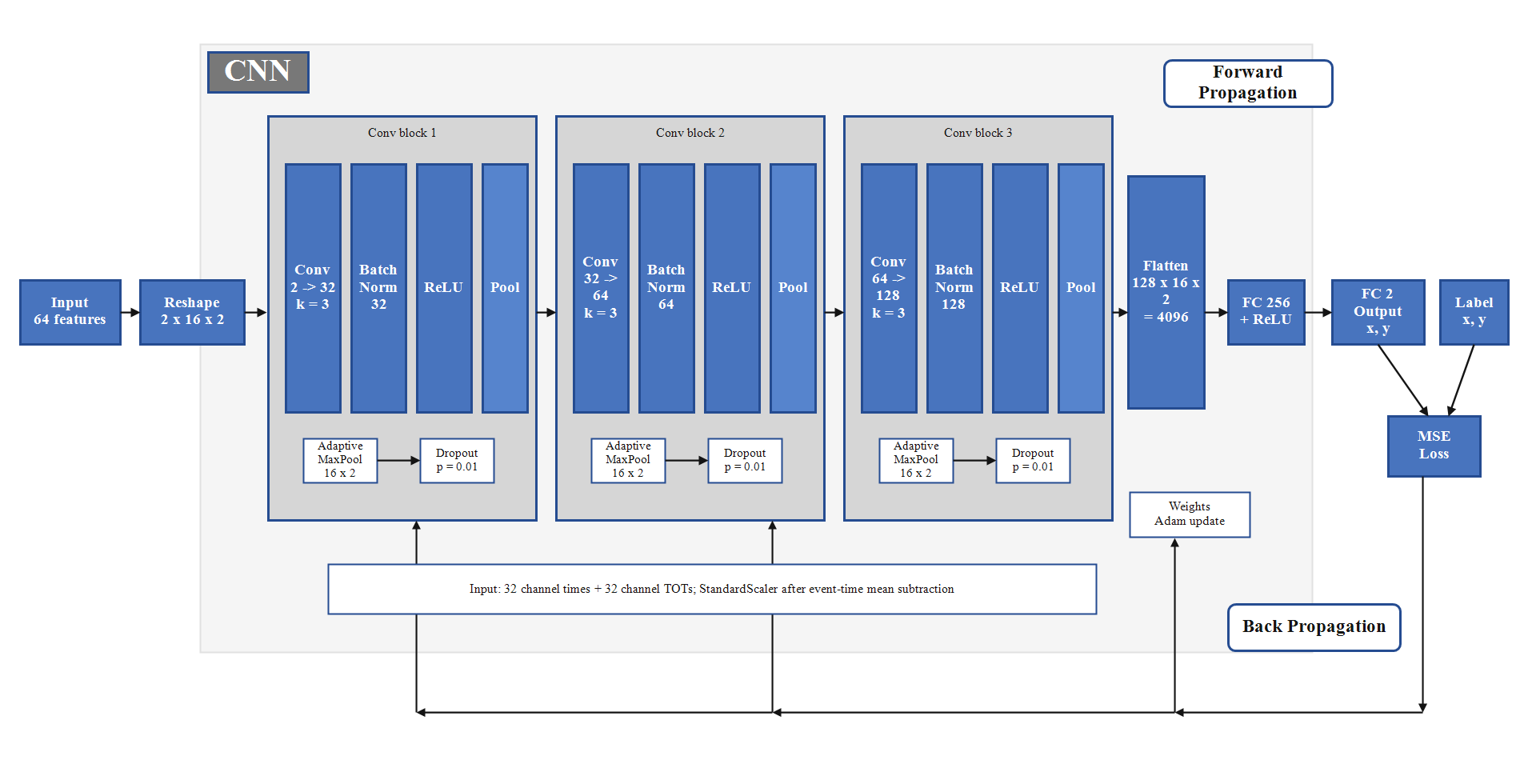}
  \caption{CNN architecture used for position reconstruction of the T0 detector.}
  \label{fig:cnn_arch}
\end{figure}

The network architecture is shown in Fig.~\ref{fig:cnn_arch}. The 64-dimensional input is first reshaped into a $2 \times 16 \times 2$ tensor. The two feature planes correspond to the relative time and event-normalized TOT information, the dimension of 16 represents the channel order of one SiPM array along the short-axis $y$ direction, and the last dimension represents the correspondence between the two arrays located at the two ends along the long-axis $x$ direction. This organization allows the convolution kernels to learn local response patterns both among neighboring channels along the short axis and between corresponding channels at the two ends.

The CNN consists of three two-dimensional convolutional modules with output channel numbers of 32, 64, and 128, respectively. Each convolution is followed by batch normalization, ReLU activation, pooling, and Dropout. The convolutional layers extract local channel-response patterns, batch normalization improves training stability, and Dropout suppresses overfitting. The convolutional features are flattened and passed to a fully connected layer with 256 neurons. The final layer outputs two continuous values corresponding to the predicted $x$ and $y$ coordinates.

\subsection{Training strategy and evaluation metrics}

During training, a subset of the labeled samples is first reserved as an independent test set. This test set is not used for network-parameter training, learning-rate adjustment, early stopping, or model selection. The remaining samples are divided into training and validation sets with a ratio of 8:2. The training set is used to update the network weights, while the validation set is used to monitor generalization error and select the best model during training. Feature-standardization parameters are calculated only from the training set and are then applied to the validation and independent test sets to avoid information leakage.

The network is trained using the Adam optimizer with an initial learning rate of 0.001 and weight decay to constrain model complexity. A learning-rate scheduling strategy is used to reduce the learning rate when the validation loss reaches a plateau, and early stopping is applied to avoid overfitting during long training. To enhance the learning of continuous position responses, mixup data augmentation can be used during training, in which the inputs and labels of different events are linearly combined to provide a smoother response between scan positions.

The final positioning performance is evaluated on the independent test set. For each test event, the residuals are defined as $\Delta x=x_{\mathrm{pred}}-x_{\mathrm{true}}$ and $\Delta y=y_{\mathrm{pred}}-y_{\mathrm{true}}$. All residual distributions presented for both simulation and experiment are obtained from data that are not included in the training or validation sets. A double-Gaussian function is used in the residual plots to illustrate the shape of the core region and tail events, but the position resolution is not taken from the width of a single Gaussian component.

In the following simulation and experimental results, the standard deviation (Std Dev) of the residual distribution on the independent test set is used as the position resolution. This metric is calculated using all test events and reflects the contributions of both core and tail events to the overall position reconstruction performance.

\section{Simulation study}
\label{sec:simulation}

\subsection{Geant4 simulation and waveform digitization}

\begin{figure}
  \centering
  \includegraphics[width=0.85\linewidth]{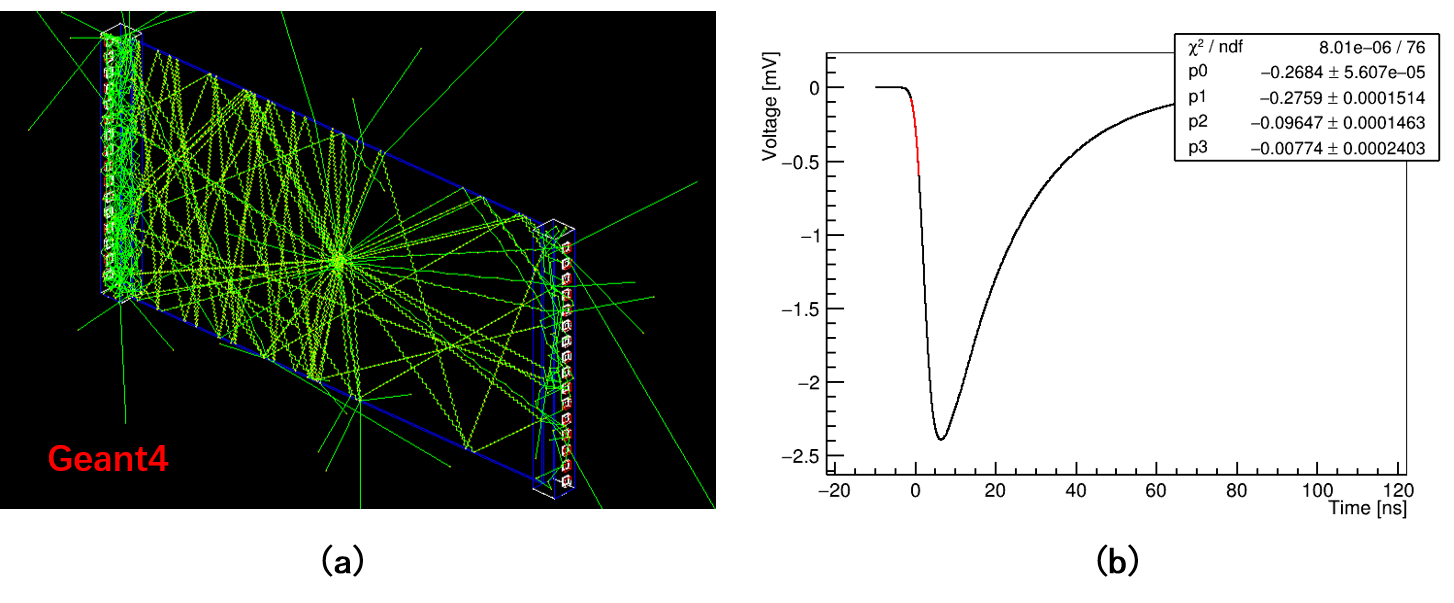}
  \caption{(a)Geant4 simulation schematic of the T0 detector.(b)SiPM waveform digitization and feature extraction.}
  \label{fig:g4}
\end{figure}

To study the algorithm performance under known true incident positions, a Geant4\cite{agostinelli2003, allison2006} model of the T0 detector was developed. In the simulation, particles are incident normally on the scintillator surface, deposit energy in the medium, and generate scintillation photons. The photons propagate and reflect inside the scintillator and are finally received by the SiPM arrays. Different incident positions change the number and arrival-time distribution of photons received by each channel, so the multi-channel optical response contains two-dimensional position information. A schematic view of the Geant4 simulation is shown in Fig.~\ref{fig:g4} (a).

Geant4 mainly provides energy deposition, photon production, and photon-arrival information at the SiPMs, whereas the experimental readout chain finally provides the over-threshold time and TOT after NINO discrimination. To make the simulation samples consistent with the experimental input format, the simulated data are processed through SiPM response and electronics digitization, rather than using photon counts or ideal arrival times directly as network inputs.

Specifically, for each event and each SiPM channel, a simulated SiPM voltage pulse is first constructed from the photon-arrival time sequence at that channel. The signal digitization of the detector is accomplished by reconstructing the output waveform of the SiPM based on the simulated number of photoelectrons (NPE) detected by the SiPM and the arrival time of each individual photoelectron. The single photon–electron (SPE) response waveform $v_{i}(t)$ can be simulated using an experimentally measured SPE signal of the SiPM combined with polynomial fitting. The output signal of the SiPM is the superposition of all the SPE signals in one event, which is given\cite{jones2000cdf}:

\begin{equation}
V(t)=\sum_{i=1}^{NPE}v_i\{t-t_{transit}-(t_0)_i\}
\label{equation:V(t)}
\end{equation}

where  $t_{transit}$ represents the transit time of a single photon within the SiPM, which is obtained by sampling from a Gaussian distribution with a mean of 0 and a standard deviation of 20 ps (in the time resolution analysis, the specific value of $t_{transit}$ does not affect the outcome). $t_0$ indicates the arrival time of a single photoelectron at the SiPM. Fig.~\ref{fig:g4} (b) shows the SiPM pulse shape used for digitization.This procedure converts photon arrival times and photon counts into simulated waveforms corresponding to the experimental electronics readout.


A fixed threshold $V_{\mathrm{thr}}$ is then applied to the simulated waveform to emulate the discrimination process of the NINO front-end electronics. The leading-edge over-threshold time is obtained when the waveform first crosses the threshold, and the trailing-edge time is obtained when the waveform falls below the threshold. Their difference defines the TOT:
\begin{equation}
t_i^{\mathrm{lead}}=\min\{t \mid V_i(t) \ \mathrm{crosses} \ V_{\mathrm{thr}}\},
\qquad
\mathrm{TOT}_i=t_i^{\mathrm{trail}}-t_i^{\mathrm{lead}} .
\label{eq:tot}
\end{equation}
In the conversion of the simulation samples, the channel timing feature is directly taken as the leading-edge over-threshold time obtained by threshold discrimination, i.e. $t_i=t_i^{\mathrm{lead}}$. The TOT feature is the over-threshold width $\mathrm{TOT}_i$ under the same threshold. The 32 channel times are then subjected to event-mean subtraction, and the 32 channel TOT values are normalized by the event-wise trimmed mean. This forms the 64-dimensional CNN input composed of relative time and event-normalized TOT features in the same way as for the experimental data.

The purpose of this digitization step is to convert the optical response in Geant4 into the same type of features as those provided by the actual FEE+TDC readout. With this processing, the simulation samples include the effects of light propagation, SiPM pulse formation, threshold discrimination, and over-threshold pulse width, allowing the network trained on simulation data to learn multi-channel response patterns closer to the experimental readout.

\subsection{Simulation data set}

The simulation data set is mainly generated using a uniform planar incident distribution. The incident positions are randomly distributed over the active area of the T0 detector, so the simulation labels cover the two-dimensional response variation across the detector plane. For each event, the time and TOT features of the 32 channels are recorded, and the true two-dimensional incident coordinates specified in Geant4 are used as labels. To keep the processing consistent with the experimental data, the simulation samples are also processed by event-mean time subtraction, event-wise trimmed-mean TOT normalization, two-side channel interleaving, and feature standardization.

This uniformly distributed planar sample is used to evaluate the CNN reconstruction capability under ideal known-position and high-density spatial-sampling conditions. In contrast, the experiment cannot provide truly continuous incident-position labels, and data can only be collected at finite scan positions using the positioning plate. Therefore, an additional simulation sample with discrete positions matching the hole positions of the laser-scanning plate is generated to evaluate the effect of sparse scanning labels on the reconstruction result. The hole-position setting is introduced in Section~\ref{sec:exp_setup}, and the corresponding simulation reference is discussed together with the experimental scan results in Section~\ref{sec:scan_sim}.

For the uniformly distributed planar sample, the training, validation, and test samples are separated by event to evaluate the reconstruction performance for events not used in training or model selection. For the scanning-geometry simulation sample, the data are organized according to the laser-scanning hole positions described in Section~\ref{sec:exp_setup}, making the spatial sampling comparable between simulation and experiment.

\subsection{Simulation results}

\begin{figure}
  \centering
  \includegraphics[width=0.95\linewidth]{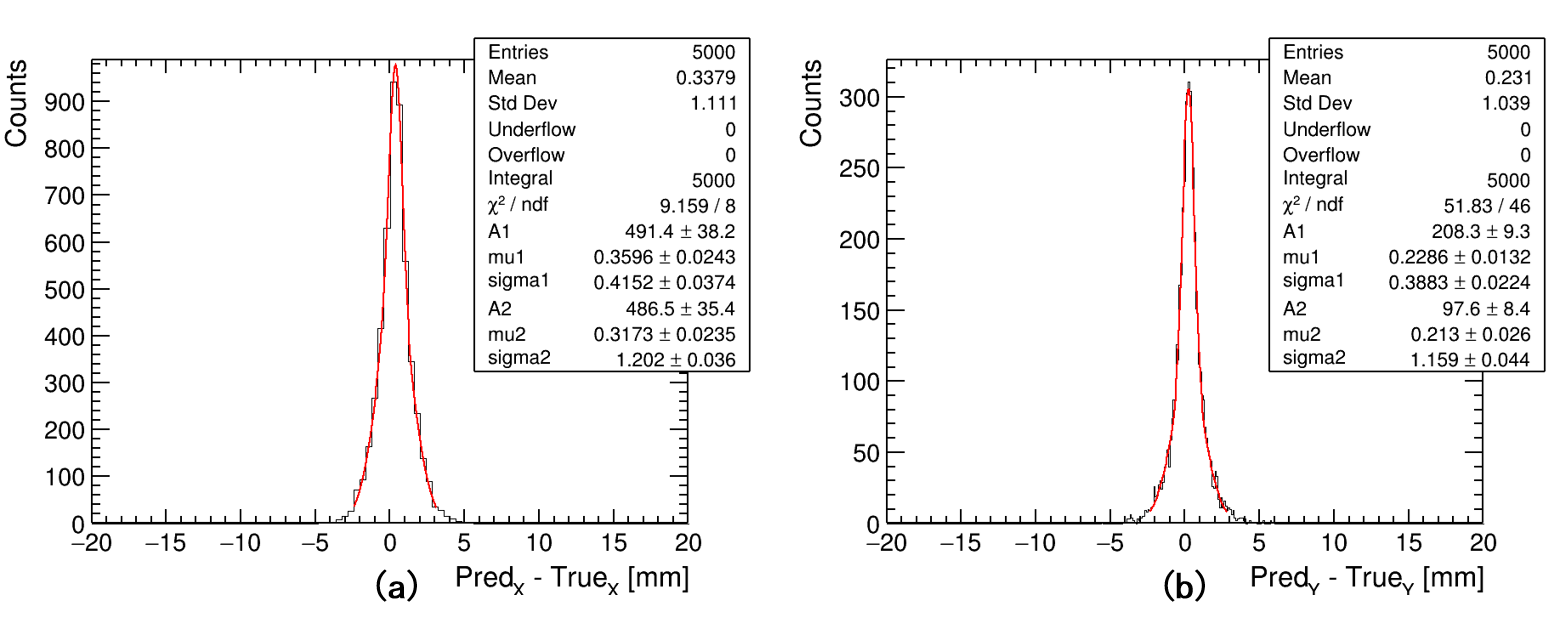}
  \caption{(a) Position residual distribution in the $x$ direction for the uniformly distributed planar simulation data. (b) Position residual distribution in the $y$ direction for the uniformly distributed planar simulation data.}
  \label{fig:sim_uniform}
\end{figure}

Figs.~\ref{fig:sim_uniform}(a) and \ref{fig:sim_uniform}(b) show the residual distributions in the $x$ direction (long axis) and $y$ direction (short axis), respectively, for the uniformly distributed planar simulation data. A double-Gaussian fit is used to illustrate the core and tail structure of the residual distributions. Using the standard deviation of the residuals as the position-resolution metric, the resolutions are approximately 1.111 mm in $x$ and 1.039 mm in $y$.

The uniformly distributed planar simulation result shows that the CNN can learn the nonlinear relationship between the incident position and the detector output from multi-channel time and TOT responses. In this sample, the incident-position labels are continuous and spatially dense, so the result is closer to the algorithm performance under ideal continuous-position learning conditions. To avoid interrupting the description before introducing the experimental scanning geometry, the simulation reference using the same scanning positions as the experiment is presented in Section~\ref{sec:scan_sim}.

\vspace{0.0cm}
\section{Experimental performance}
\label{sec:experiment}

\vspace{0.0cm}
\subsection{Laser scanning setup and data set}
\label{sec:exp_setup}

To validate the feasibility of the CNN-based position reconstruction method under real readout conditions, a pulsed-laser scanning setup was built. The general idea is similar to a fixed-point laser readout test, in which a controllable light source excites the scintillator and the SiPM array output is recorded. The difference is that the present experiment does not evaluate the electronics response at only one fixed point, but scans multiple positions over the active area of the detector to construct a data set with known two-dimensional position labels.

\begin{figure}
  \centering
  \includegraphics[width=0.95\linewidth]{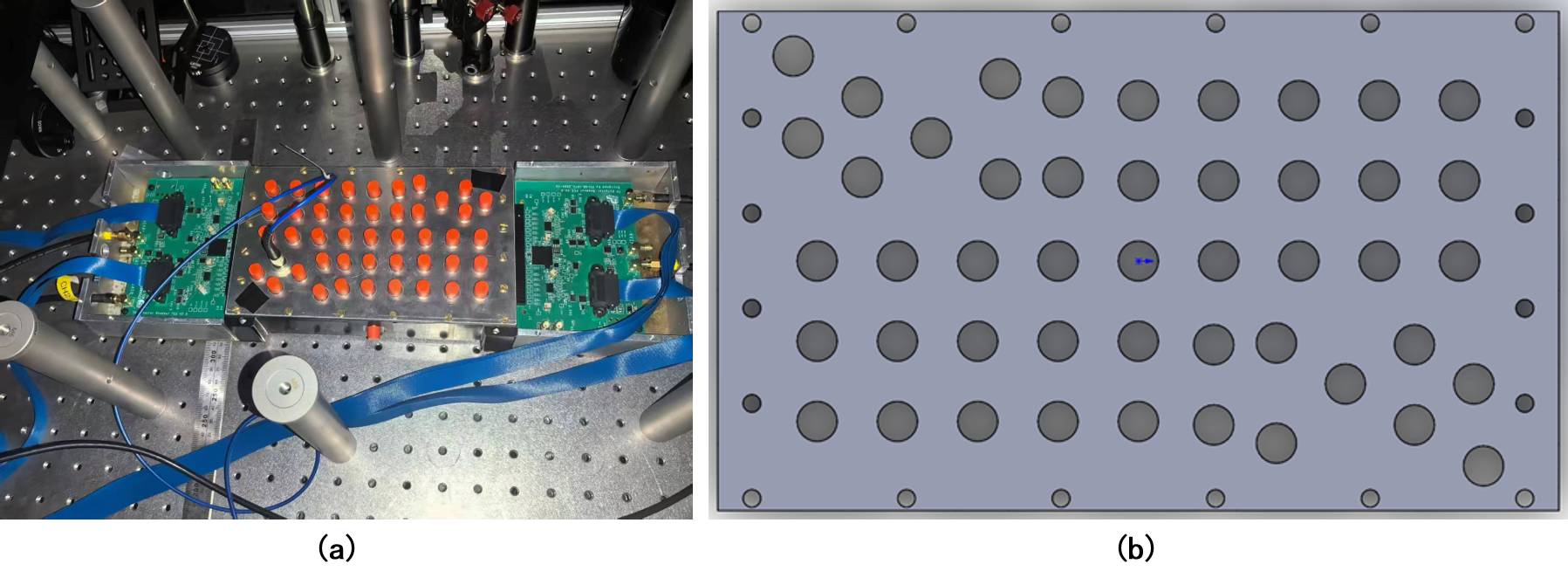}
  \caption{(a)Laser-scanning experimental platform for the T0 detector.(b)Hole-position distribution used in the laser-scanning experiment.}
  \label{fig:exp_platform}
\end{figure}

In the experiment, the T0 detector is installed on an optical table, and the pulsed laser is injected into specified positions of the detector through an optical fiber. A photograph of the experimental platform is shown in Fig.~\ref{fig:exp_platform}(a), where the T0 detector body, the readout boards on both sides, and the positioning structure mounted on the optical table can be seen. The platform provides reproducible geometrical constraints for the laser scan.

A scanning positioning plate with holes is placed above the detector, and each hole corresponds to a known two-dimensional coordinate. During the experiment, the holes are selected sequentially for laser illumination, and multiple events are recorded at each scan position. In this way, the hole coordinates provide the supervised-learning labels, while the 32-channel readout responses provide the network inputs. The hole distribution of the scanning plate is shown in Fig.~\ref{fig:exp_platform}(b).

Fig.~\ref{fig:exp_platform}(b) shows the hole-position distribution of the scanning plate used in the experiment. The hole coordinates are used as supervised-learning labels, and the multi-channel responses acquired point by point are used as network inputs. Compared with an evaluation based only on simulation samples, this data set includes real SiPM response, channel-to-channel differences, front-end electronics effects, and practical alignment errors, and therefore more directly reflects the algorithm performance in the experimental environment.

\subsection{Simulation reference under the laser-scanning geometry}
\label{sec:scan_sim}

Based on the hole positions of the scanning plate, an additional simulation sample with the same incident-position setting as the experiment was generated. This reference sample was still constructed using the Geant4 and waveform-digitization procedure, but the incident positions were no longer sampled from a uniform planar distribution. Instead, they were fixed to the experimental scanning hole positions. This allows the effect of finite scan positions on supervised learning to be studied without introducing real optical-alignment, mechanical-installation, or channel-nonuniformity effects.

\begin{figure}
  \centering
  \includegraphics[width=0.95\linewidth]{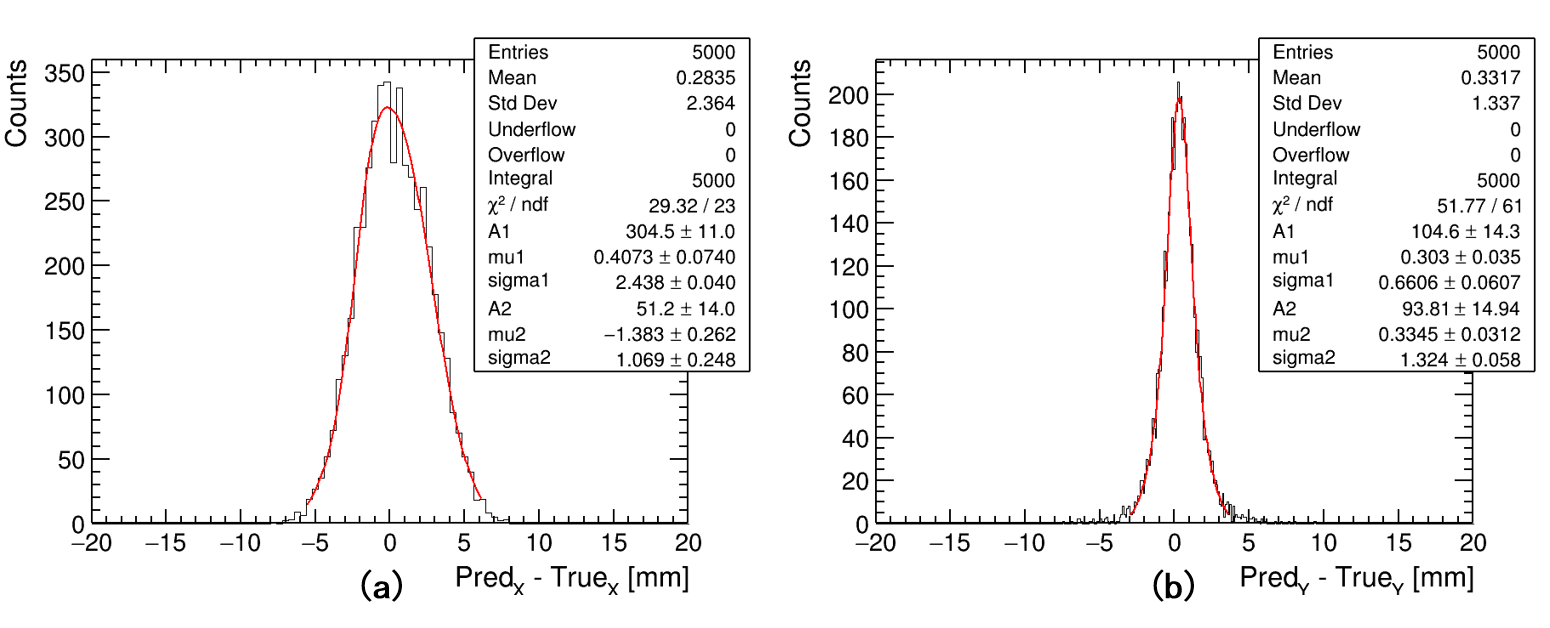}
  \caption{(a)Position residual distribution in the $x$ direction for the simulation data generated with the same incident-position setting as the experiment.(b)Position residual distribution in the $y$ direction for the simulation data generated with the same incident-position setting as the experiment}
  \label{fig:sim_scan}
\end{figure}

Figs.~\ref{fig:sim_scan}(a)and \ref{fig:sim_scan}(b) show the residual distributions in the $x$ direction (long axis) and $y$ direction (short axis), respectively, for the scanning-geometry simulation sample. The corresponding position resolutions are approximately 2.364 mm and 1.337 mm, which are clearly worse than the 1.111 mm and 1.039 mm obtained from the uniformly distributed planar simulation. This indicates that, under the present data organization, changing the spatial labels from a continuous distribution to discrete scan positions broadens the reconstruction result, and that the density of the finite scan positions is an important factor affecting the experimental validation.

Thus, the uniformly distributed planar simulation in Section~\ref{sec:simulation} represents the algorithm capability under relatively ideal conditions, while the scanning-geometry simulation in this section is closer to the spatial-sampling scheme of the experimental data. Combining these two results helps separate the intrinsic reconstruction capability of the algorithm from the effect of scan-label density.

\subsection{Experimental preprocessing}

The preprocessing of the experimental data follows the same procedure as that used for the simulation data. First, the over-threshold time and TOT information of the 32 channels are extracted for each event. The time features are then corrected by subtracting the event mean time to establish an event-internal time reference for the T0 detector. The TOT features are normalized by the event-wise trimmed mean to suppress event-to-event variations in the overall signal scale while retaining the relative response distribution among channels. The features are further rearranged according to the correspondence between the two detector sides and standardized using the training-set statistics. Finally, each event forms a 64-dimensional input vector with a two-dimensional scan-hole coordinate label.

Because the experiment includes laser-coupling stability, hole-position machining and installation errors, SiPM gain differences, and electronics-threshold variations, the response distribution of the experimental data is usually more complex than that of the simulation data. The advantage of the CNN is that it does not require an explicit analytical model for each non-ideal effect, but instead learns the mapping between multi-channel responses and true positions directly from the training samples.

\subsection{Experimental results}

\begin{figure}
  \centering
  \includegraphics[width=0.95\linewidth]{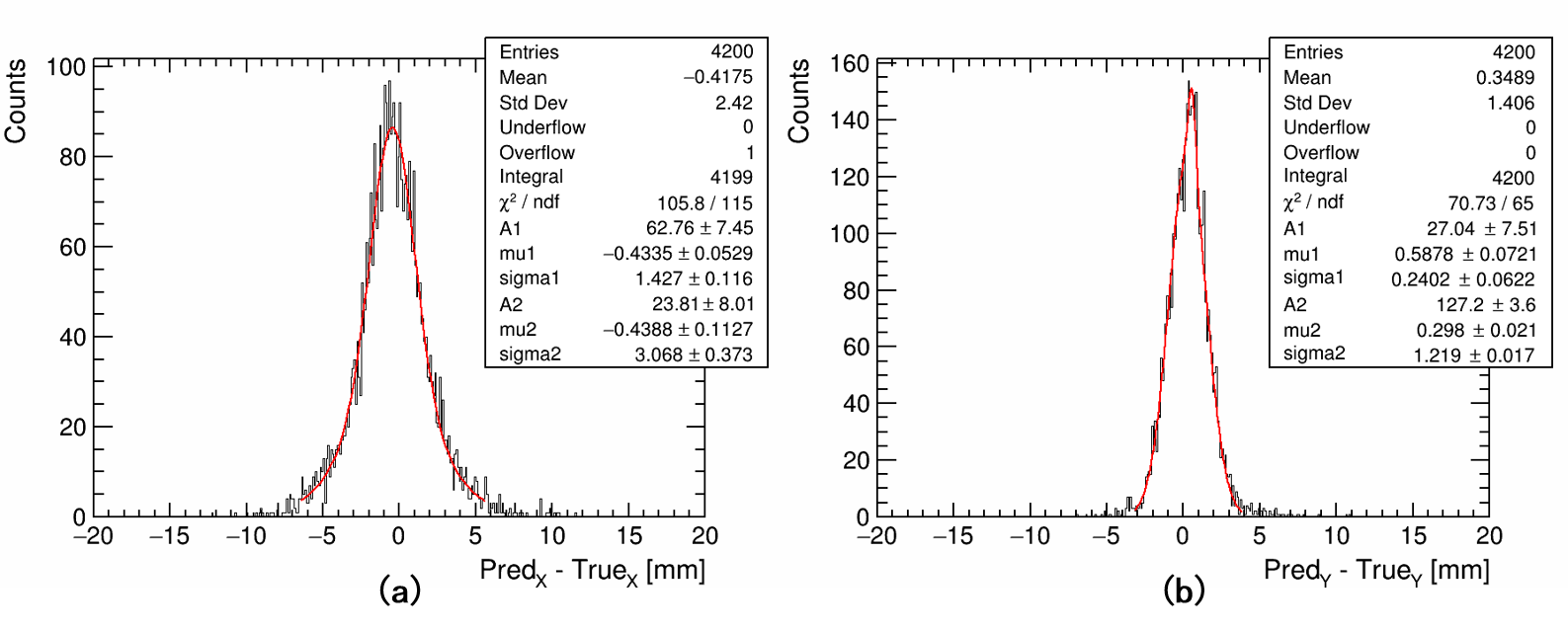}
  \caption{(a)Position residual distribution in the $x$ direction for the experimental scanning data.(b)Position residual distribution in the $y$ direction for the experimental scanning data}
  \label{fig:exp}
\end{figure}

Figs.~\ref{fig:exp}(a) and \ref{fig:exp}(b) show the residual distributions in the $x$ direction (long axis) and $y$ direction (short axis), respectively, for the experimental scanning data. Compared with the uniformly distributed planar simulation result, the experimental residual distributions are broader and show a certain position offset, which is related to optical-path alignment, scan-hole position errors, channel-response nonuniformity, and electronics noise in the real system. The experimental data give position resolutions of approximately 2.420 mm in $x$ and 1.406 mm in $y$.

The experimental results show that the CNN method still achieves millimeter-scale two-dimensional positioning capability with real scanning data. Although the experimental resolution is worse than that of the uniformly distributed planar simulation, it is close to the scanning-geometry simulation result in Section~\ref{sec:scan_sim}. This indicates that the spatial sampling density of the finite scan positions is one of the important factors limiting the present experimental resolution. At the same time, laser-coupling stability, hole-position machining and installation errors, channel-response nonuniformity, and electronics-threshold differences may also contribute to the residual broadening. Further improvements can be expected by increasing the number of scan positions, improving the laser-injection position accuracy, refining channel-to-channel gain calibration, and optimizing the network architecture.

\section{Beam-spot observation}
\label{sec:beamspot}

\begin{figure}
  \centering
  \includegraphics[width=0.75\linewidth]{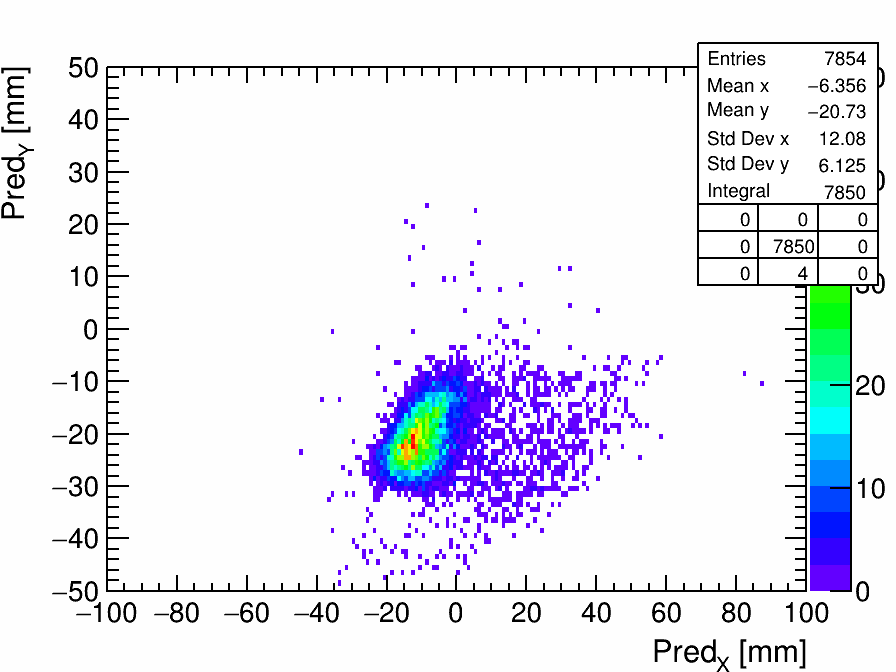}
  \caption{Two-dimensional beam-spot distribution reconstructed from the CEE zero-field beam data using the CNN-based position reconstruction method.}
  \label{fig:beamspot}
\end{figure}

After validation with simulation and scanning-experiment data, the trained CNN model was applied to beam data collected during the CEE zero-field beam experiment at the Institute of Modern Physics, Lanzhou, in January 2026. Event-by-event two-dimensional positions were reconstructed for the beam events. By filling the predicted coordinates of many events into a two-dimensional histogram, the beam-spot distribution in the T0 detector plane can be obtained. The beam-spot observation result for this beam data set is shown in Fig.~\ref{fig:beamspot}.

The statistics shown in the figure indicate that this beam data sample contains approximately 7854 events. The mean reconstructed beam-spot position is approximately $x=-6.356$ mm and $y=-20.73$ mm, with standard deviations of $\sigma_x=12.08$ mm and $\sigma_y=6.125$ mm. This result shows that the multi-channel response of the T0 detector can be used not only for event start-time measurement, but also for online or offline analysis of the beam-position distribution in beam experiments. In practical applications, the absolute beam-spot coordinates should be further calibrated by combining the detector installation coordinate system with the beam-line coordinate system.

\vspace{0.0cm}
\section{Summary}
\label{sec:summary}

A CNN-based position reconstruction algorithm for the T0 detector in the HIRFL-CSR external-target experiment has been studied. The method combines the relative timing and event-normalized TOT information from 32 SiPM channels into a 64-dimensional input vector and reshapes it into a tensor according to the detector-channel layout. The CNN directly outputs the two-dimensional incident coordinates through three convolutional modules and fully connected regression layers, avoiding the difficulty of constructing an analytical reconstruction model under complex light-transport and electronics-response conditions.

The Geant4 simulation and waveform-digitization study shows that the CNN achieves millimeter-scale position resolution for uniformly distributed planar simulation data, with resolutions of approximately 1.111 mm in $x$ and 1.039 mm in $y$. For simulation samples generated with the same incident-position setting as the experimental scan, the position resolutions are approximately 2.364 mm in $x$ and 1.337 mm in $y$. The pulsed-laser scanning data with known hole positions further validate the feasibility of the method under real readout conditions, giving experimental position resolutions of approximately 2.420 mm in $x$ and 1.406 mm in $y$.

The simulation result with the same scanning geometry as the experiment is close to the experimental result, indicating that the spatial sampling density of the finite scan positions is one of the important factors limiting the present experimental resolution. Increasing the number of scan positions, improving the scan-position accuracy, and optimizing the scanning grid will be important for improving the experimental positioning performance. Finally, the trained model was applied to beam data collected during the CEE zero-field beam experiment at the Institute of Modern Physics, Lanzhou, in January 2026, and an event-by-event two-dimensional beam-spot distribution was obtained. The results suggest that the T0 detector combined with a CNN algorithm can provide additional information for beam-position monitoring and experimental-operation diagnostics while retaining its role as an event start-time detector. Future work will include further experimental-parameter calibration, expansion of the scanning data set, optimization of the channel-calibration procedure, and studies of the long-term stability of the model in beam-experiment data.

\printcredits

\bibliographystyle{unsrtnat}
\bibliography{references}

@article{CSRext,
  author  = {L\"u, L. M. and Yi, H. and Xiao, Z. G. and others},
  title   = {Conceptual Design of the HIRFL-CSR External-Target Experiment},
  journal = {Sci. China Phys. Mech. Astron.},
  volume  = {60},
  pages   = {012021},
  year    = {2017}
}

@article{Luo2017NST,
  author  = {Luo, X. and Xu, N.},
  title   = {Search for the QCD critical point with fluctuations of conserved quantities in relativistic heavy-ion collisions at RHIC: an overview},
  journal = {Nucl. Sci. Tech.},
  volume  = {28},
  pages   = {112},
  year    = {2017}
}

@article{Hu2023T0,
  author  = {Hu, D. and Wang, X. and Shao, M. and others},
  title   = {Design and Performance Testing of a T0 Detector for the CSR External-target Experiment},
  journal = {Nucl. Instrum. Methods Phys. Res. A},
  volume   = {1057},
  pages    = {168773},
  year     = {2023}
}

@article{Alimena2020,
  author  = {Alimena, J. and Iiyama, Y. and Kieseler, J.},
  title   = {Fast Convolutional Neural Networks for Identifying Long-Lived Particles in a High-Granularity Calorimeter},
  journal = {J. Instrum.},
  volume  = {15},
  number  = {12},
  pages   = {P12006},
  year    = {2020}
}

@article{Polson2022EnergyCNN,
  author  = {Polson, L. and Kurchaninov, L. and Lefebvre, M.},
  title   = {Energy Reconstruction in a Liquid Argon Calorimeter Cell Using Convolutional Neural Networks},
  journal = {J. Instrum.},
  volume  = {17},
  pages   = {P01002},
  year    = {2022}
}

@techreport{ATLAS2019CNN,
  author      = {{ATLAS Collaboration}},
  title       = {Convolutional Neural Networks with Event Images for Pileup Mitigation with the ATLAS Detector},
  institution = {CERN},
  number      = {ATL-PHYS-PUB-2019-028},
  year        = {2019}
}

@article{MicroBooNE2021CNNPID,
  author  = {{MicroBooNE Collaboration}},
  title   = {Convolutional Neural Network for Multiple Particle Identification in the MicroBooNE Liquid Argon Time Projection Chamber},
  journal = {Phys. Rev. D},
  volume  = {103},
  pages   = {092003},
  year    = {2021}
}

@misc{EJ200,
  author       = {{Eljen Technology}},
  title        = {EJ-200 Plastic Scintillator},
  howpublished = {\url{https://eljentechnology.com/products/plastic-scintillators/ej-200-ej-204-ej-208-ej-212}}
}

@misc{S13360,
  author       = {{Hamamatsu Photonics K.K.}},
  title        = {S13360-3025PE MPPC Array},
  howpublished = {\url{https://www.hamamatsu.com/eu/en/product/optical-sensors/mppc/mppc_mppc-array/S13360-3025PE.html}}
}

@inproceedings{Cao2022ICETIS,
  author    = {Cao, J. and Lu, J. and Zhao, L. and others},
  title     = {Prototype Design of the Time Measurement Circuit for SiPM-based T0 Detector of CEE},
  booktitle = {Proceedings of the 7th International Conference on Electronic Technology and Information Science (ICETIS)},
  pages      = {1--6},
  year       = {2022}
}

@book{Goodfellow2016,
  author    = {Goodfellow, Ian and Bengio, Yoshua and Courville, Aaron},
  title     = {Deep Learning},
  publisher = {MIT Press},
  address   = {Cambridge, MA},
  year      = {2016}
}

@inproceedings{batchnorm,
  author    = {Ioffe, Sergey and Szegedy, Christian},
  title     = {Batch Normalization: Accelerating Deep Network Training by Reducing Internal Covariate Shift},
  booktitle = {Proceedings of the 32nd International Conference on Machine Learning},
  series     = {Proceedings of Machine Learning Research},
  volume     = {37},
  pages      = {448--456},
  year       = {2015}
}

@article{Agostinelli2003,
  author  = {Agostinelli, S. and Allison, J. and Amako, K. and others},
  title   = {Geant4--a simulation toolkit},
  journal = {Nucl. Instrum. Methods Phys. Res. A},
  volume  = {506},
  pages   = {250--303},
  year    = {2003}
}

@article{Allison2006,
  author  = {Allison, J. and Amako, K. and Apostolakis, J. and others},
  title   = {Geant4 Developments and Applications},
  journal = {IEEE Trans. Nucl. Sci.},
  volume  = {53},
  pages   = {270--278},
  year    = {2006}
}

@techreport{Jones2000CDF,
  author      = {Jones, M.},
  title       = {PMT Base Design for the CDF-II Time-of-Flight System},
  institution = {Fermilab},
  number      = {CDF Note 5358},
  year        = {2000}
}

\end{document}